 \documentclass[aps,prl,twocolumn,groupedaddress,amsmath,amssymb,showpacs]{revtex4}
\usepackage{graphicx}
\usepackage{amsmath}
\usepackage{amssymb}
\usepackage{mathrsfs}
\usepackage{amsfonts}
\usepackage{dcolumn}
\usepackage{bm}

\newcommand{\vvr}{\mathbf{r}}

\newcommand{\vk}{\mathbf{k}}

\newcommand{\vE}{\mathbf{E}}
\newcommand{\vB}{\mathbf{B}}

\newcommand{\hvu}{\hat{\mathbf{u}}}
\newcommand{\hvn}{\hat{\mathbf{n}}}

\newcommand{\hvx}{\hat{\mathbf{x}}}
\newcommand{\hvy}{\hat{\mathbf{y}}}
\newcommand{\hvz}{\hat{\mathbf{z}}}

\newcommand{\bp}{\bm{p}}
\newcommand{\bj}{\bm{j}}
\newcommand{\bP}{\bm{P}}
\newcommand{\bJ}{\bm{J}}
\providecommand{\abs}[1]{\lvert#1\rvert}

\newcommand{\lbar}{{\lambda \negthickspace \! \text{\rule[5.5pt]{0.135cm}{.2pt}} \, }}
\newcommand{\di}{\mathrm{d}}

\begin{document}
\title{Geometric Spin Hall Effect of Light}
\author{Andrea Aiello$^{1,*}$}
\author{Christoph Marquardt$^{1,2}$}
\author{Gerd Leuchs$^{1,2}$}
\affiliation{${^1}$Max Planck Institute for the Science of Light, G\"{u}nter-Scharowsky-Str. 1/Bau 24, 91058 Erlangen, Germany}
\affiliation{${^2}$Institute for Optics, Information and Photonics, University Erlangen-Nuernberg,
Staudtstr. 7/B2, 91058 Erlangen, Germany.}
\affiliation{$^*$Corresponding author: Andrea.Aiello@mpl.mpg.de}
\begin{abstract}
We describe a novel phenomenon occurring when a  polarized Gaussian beam of light is observed in a Cartesian reference frame whose axes are not parallel to the direction of propagation  of the beam.
Such phenomenon amounts to an  intriguing spin-dependent shift of the position of the center of the  beam, with manners akin to the spin Hall effect of light.
 We demonstrate that this effect is  unavoidable  when the light beam possesses a nonzero transverse  angular momentum.
\end{abstract}
\maketitle
{\flushleft\emph{\hspace{.3cm}Introduction.$\,$}\rule[1.8pt]{0.3cm}{.4pt}}
Angular momentum (AM) of light beams is a topic that has recently attracted the attention of many researcher, from both the classical and quantum optics communities \cite{BarnettBook}.
However, all the  studies produced up to now  deal exclusively with the \emph{longitudinal} component $J_z$ of the AM. Surprisingly enough,  not attention at all was devoted to the study of the  transverse components of the AM of a light beam.

With this manuscript we begin a systematic investigation of the properties of the  components of the AM perpendicular to the propagation axis of the beam. As it will be shown below, these components may be responsible for some intriguing and counterintuitive effects, as the ``geometric spin Hall effect of light''  (geometric SHEL, for short) presented herewith. SHEL is an interesting phenomenon occurring when a light beam impinges upon a planar interface separating two different media \cite{OnodaEtAlPRL,BliokhPRL}. In practice, when a linearly polarized beam of light is reflected or transmitted by such interface, it splits into its two left/right circularly polarized components. This split occurs in a direction perpendicular to the plane of incidence: It has been recently observed for a beam transmitted across an air-glass interface \cite{HostenandKwiat}.
The occurrence of a similar left/right shift, affecting polarized light beams propagating along curved trajectories, has also been predicted and observed \cite{BliokhNP}.

In this manuscript we describe a third type of left/right shift that occurs when a linearly polarized Gaussian beam  is observed by means of a detection system whose optical axis is not parallel to the propagation direction  of the beam.
We show that this shift is inherently connected with the existence of a transverse part of the angular momentum  of the beam.
{\flushleft\emph{\hspace{.3cm}Linear and angular momentum of a light beam.$\,$}\rule[1.8pt]{0.3cm}{.4pt}}
A monochromatic electromagnetic beam in vacuo possesses  a time-averaged \emph{linear} ($\bp$) and an \emph{angular} ($\bj$) momentum density equal to
\begin{subequations}\label{momenta}
\begin{align}
\bp(\vvr) & =  \epsilon_0  \mathrm{Re} \left[ \vE(\vvr) \times  \vB^*(\vvr)\right], \label{linear}\\
\bj(\vvr) & =  \vvr \times \bp(\vvr) , \label{angular}
\end{align}
\end{subequations}
where $\mathrm{Re}  \left[ \vE(\vvr)e^{- i \omega t} \right]$ and $\mathrm{Re} \left[ \vB(\vvr)e^{- i \omega t}\right]$ are the  time-harmonic electric and magnetic fields of the beam, respectively \cite{MandelBook}.
The linear momentum density $\bp$ is equal to $1/c^2$ the Poynting vector.
In the quantum theory of light, a  polarized photon of energy $\hbar \omega$ propagating in the $z$ direction has a $z$ component of spin angular momentum (helicity) of $\sigma \hbar$, where $\sigma = \pm 1$ for a circularly polarized photon and $\sigma = 0$ for a linearly polarized photon \cite{HausandPan}. Analogously, classical electrodynamics establishes that for a paraxial beam of light propagating in the $z$ direction,  the ratio of spin angular momentum density flux along $z$ to energy density flux along $z$ is
\begin{align} \label{ratioFlux}
\frac{J_z^\mathrm{spin}}{c P_z}  = \frac{\sigma}{\omega},
\end{align}
where $J_z = \hvz \cdot \bJ(z)$, $P_z = \hvz \cdot  \bP(z)$, and  $\bP(z)$ and $\bJ(z)$ are the linear and angular momentum of the beam per unit length obtained by integrating $\bp$ and $\bj$ over the $x$-$y$ plane, respectively \cite{HausandPan,Barnett&Allen}.
At any plane orthogonal to the $z$ direction, the  intensity of the beam $I(\vvr) = c^2 p_z(\vvr) $, can be regarded as the \emph{spatial} probability  distribution of the transverse coordinate vector $\vvr_\perp = \hvx x + \hvy y $. The mean value $\langle \vvr_\perp \rangle = \hvx \langle x \rangle + \hvy \langle y \rangle$ with respect to the distribution $I(\vvr)$ is
\begin{align} \label{centroid}
\langle \vvr_\perp \rangle= \frac{\displaystyle{ \iint \vvr_\perp p_z( \vvr)\, \di x \di y}}{\displaystyle{\iint p_z( \vvr) \, \di x \di y }},
\end{align}
and it determines the centroid (or barycenter) of the beam.
Note that the denominator of Eq. (\ref{centroid}) is, by definition, equal to $P_z$. From this fact and
 Eq. (\ref{angular}), it immediately follows that
\begin{subequations}\label{Jperp}
\begin{align}
J_x & =   \left\langle y \right\rangle P_z - z P_y,   \label{Jx} \\
 J_y & =   z P_x  -  \left\langle x \right\rangle P_z,   \label{Jy}
\end{align}
\end{subequations}
that, in the plane $z=0$,  reduce to
\begin{align} \label{Jperp0}
{J_x}/{P_z} = \left\langle y \right\rangle, \qquad {J_y}/{P_z}  = - \left\langle x \right\rangle.
\end{align}
This remarkably simple result shows that  the centroid of a beam with a nonzero \emph{transverse} angular momentum per unit length $\bJ_\perp = \hvx J_x + \hvy J_y$ and propagating in the $z$ direction, is displaced with respect to the propagation axis $z$  in a direction orthogonal to $\bJ_\perp$:
\begin{align} \label{ratioJ}
\left. \langle \vvr_\perp \rangle \cdot \bJ_\perp \right|_{z=0}  = 0.
\end{align}
 Equation (\ref{Jperp}) automatically furnishes a simple recipe to actually \emph{measure} the transverse angular momentum of the beam, by measuring the position of its centroid in $z=0$ \cite{TrepsPointer}.
 \vspace{-0.22cm}
{\flushleft\emph{\hspace{.3cm}Paraxial beams.$\,$}\rule[1.8pt]{0.3cm}{.4pt}}
We will calculate the linear and angular momentum per unit length $\bP(z)$ and $\bJ(z)$, for an arbitrary paraxial beam of light. Let $f = f(\vvr)$  denote a solution of the scalar paraxial wave equation \cite{DandG}
\begin{align} \label{paraxial}
\partial_x^2 f + \partial_y^2 f + 2 i k \partial_z f   = 0,
\end{align}
where $k = \omega/c$ is the wavenumber. The electric and magnetic vector fields are expressible in terms of $f$ as:
\begin{subequations}\label{Fields}
\begin{align}
\vE(\vvr) & =  i \omega \left[ \alpha f \hvx + \beta f \hvy + i \left( \alpha \, \partial_x f + \beta  \, \partial_y f  \right)  \hvz \right],   \label{E} \\
\vB(\vvr) & =  i k \left[ -\beta f \hvx  + \alpha f \hvy  - i \left( \beta  \, \partial_x f - \alpha  \, \partial_y f \right)  \hvz  \right],   \label{B}
\end{align}
\end{subequations}
where $\hvu = \alpha \hvx + \beta \hvy$ is a complex unit vector perpendicular to $z$ that determines the polarization of the beam, and $\hvu^* \cdot \hvu = |\alpha|^2 + |\beta|^2 =1$  \cite{HausandPan}. Next, substituting Eqs. (\ref{E}) and (\ref{B}) into Eqs. (\ref{momenta}) we obtain
\begin{subequations}\label{p}
\begin{align}
k p_x = & \, - \text{Im}\left(  f  \partial_x f^* \right)  + \sigma \, \text{Re} \left( f  \partial_y f^*  \right), \label{px} \\
k p_y = & \, - \text{Im}\left(  f  \partial_y f^* \right)  - \sigma \, \text{Re} \left( f  \partial_x f^*  \right),\label{py} \\
k p_z = & \, k \abs{f }^2 , \label{pz}
\end{align}
\end{subequations}
and
\begin{subequations}\label{j}
\begin{align}
k j_x = &  k y \abs{f }^2 + z\,  \text{Im}\left(  f  \partial_y f^* \right)  + z \sigma \, \text{Re} \left( f  \partial_x f^*  \right), \label{jx} \\
k j_y = &  -k x \abs{f }^2 - z\, \text{Im}\left(  f  \partial_x f^* \right)  + z \sigma \, \text{Re} \left( f  \partial_y f^*  \right), \label{jy} \\
k j_z = & \,
-x \, \text{Im} \left( f  \partial_y f^*  \right) + y \, \text{Im} \left( f  \partial_x f^*  \right)\nonumber \\
&  - \sigma  \left[x \, \text{Re} \left( f  \partial_x f^*  \right) + y \, \text{Re} \left( f  \partial_y f^*  \right) \right], \label{jz}
\end{align}
\end{subequations}
where $\sigma = i (\alpha \beta^* - \alpha^* \beta)$ denotes the helicity of the beam, and both $\bp$ and $\bj$ are given in unit of $\hbar k$. It is easy to check that the constraint $\vvr \cdot \bj=0$ is automatically satisfied by the expressions in Eq. (\ref{j}).
We note that, in both Eqs. (\ref{p}) and (\ref{j}),  the polarization coefficients $\alpha$ and $\beta$ appears only in the combination  $\sigma = i (\alpha \beta^* - \alpha^* \beta)$, thus permitting an unambiguous identification of the spin and orbital contributions to $\bj$ \cite{BarnettFlux}.
As we are interested to the linear and orbital momenta per unit length $\bP$ and $\bJ$ respectively,  we must integrate the expressions in Eqs. (\ref{p}) and (\ref{j}) over the $x$-$y$ plane. To perform this operation we have to choose a particular function $f$. However, some general features of $\bP$ and $\bJ$ can be inferred without making such a choice. In fact, for any function $f(\vvr)$ that vanish sufficiently fast for $|\vvr_\perp| \rightarrow \infty$,   integration by parts leads to to following relations:
\begin{subequations}\label{relations}
\begin{gather}
  \smallint \mathrm{Re}\left( f \partial_x f^* \right) =  \smallint \mathrm{Re}\left( f \partial_y f^* \right) = 0, \label{rel1} \\
 \smallint \left[ x \, \text{Re} \left( f  \partial_x f^*  \right) + y \, \text{Re} \left( f  \partial_y f^*  \right) \right] =  - \smallint \abs{f }^2, \label{rel2}
\end{gather}
\end{subequations}
where $ \smallint{g}$ is a shorthand for the integration of $g$ over the $x$-$y$ plane. From Eqs. (\ref{j}) and (\ref{relations}) it follows that
%
\begin{align}
\bP = & \,\smallint f^* \left( \hvz -  i \lbar  \bm{\nabla}_{\! \! \perp}\right) f, \label{P} \\
\bJ = & \,  \smallint f^* \left[ \lbar \sigma \hvz +\vvr \times \left( \hvz - i \lbar \bm{\nabla}_\perp  \right) \right]f, \label{J}
\end{align}
%
where we used the suggestive notation $\lbar = 1/k = \lambda/(2 \pi)$ \cite{BliokhNP}. Although similar expressions for some components of $\bP$ and $\bJ$ have been given before \cite{Hugrass,vanEnk&Nienhuis,HausandPan,Barnett&Allen}, Eqs. (\ref{P}) and (\ref{J}) are novel in that they naturally embody  the  formal connection between paraxial optics and single-particle quantum mechanics. It is tempting to associate the integrand in Eq. (\ref{J}) with the total angular momentum of a single-photon wave packet with the wave vector centered in $\vk = k \hvz$ and spatially localized around $\vvr$: $\bj = \lbar \left( \vvr \times \vk  +  \sigma \vk/k \right)$, where we have neglected the corrections of order $O(\lbar \, / \abs{\vvr_\perp})$ with respect to $1$. This expression for $\bj$  in fact coincides with the corresponding one derived in the context of geometrical optics \cite{OnodaEtAlPRL,BliokhPRL}.

Further insights can be gained from  Eqs. (\ref{P}) and (\ref{J}) if we write, without loss of generality, the arbitrary function $f$ in terms of Hermite-Gaussian solutions of the paraxial wave equation: $f =  \sum_{n,m=0}^\infty  f_{nm } \psi_{nm}(\vvr)$, with $f_{n m} \in \mathbb{C}$,
and
\begin{align}\nonumber
\psi_{nm}(\vvr) = &  \sqrt{\frac{2^{1-n-m}}{\pi w^2(z) n! m!}}H_n \biggl[ \frac{\sqrt{2} \, x}{w(z)}\biggr] H_m \biggl[ \frac{\sqrt{2} \, y}{w(z)}  \biggr]  \\
& \times e^{\frac{i k}{2}\frac{x^2 + \, y^2}{z - i L}} e^{-i(n+m+1) \arctan(z/L)}. \label{HG}
\end{align}
Here $H_n\left( u \right)$ is the Hermite polynomial of degree $n$, $L = k w_0^2/2$ is the Raleigh range of a Gaussian beam with a minimum spot size $w_0$, and $w(z) = w_0 \sqrt{1 + z^2/L^2}$ is the spot size  at distance $z$ from the waist of the beam \cite{MandelBook}.
After a tedious but straightforward calculation we arrive at the following expression for $\bP$ and $\bJ$:
\begin{align}
\bP = & -i \sum_{n,m,p,q}  f_{nm}^* \bigl( \hvx \, B_{np} \delta_{mq}
 + \hvy \, \delta_{np} B_{mq} \bigr. \nonumber \\
 & \hphantom{-i \sum_{n,m,p,q}  f_{nm}^* \bigl(+} \bigl. + i \hvz \delta_{np} \delta_{mq}
 \bigr) f_{pq}, \label{bP} \\
\bJ = & \lbar \sum_{n,m,p,q} f_{nm}^*\bigl\{ \hvx \, \delta_{np} C_{mq}
 - \hvy \, C_{np} \delta_{mq} \bigr.  \nonumber \\
 &  + \hvz \left[ \sigma \delta_{np} \delta_{mq} - i  \sqrt{n q} \, \delta_{q,m+1} \delta_{n,p+1}  \right. \nonumber \\
& \hphantom{+ \hvz []}  \bigl. \left. +i  \sqrt{m p} \, \delta_{m,q+1} \delta_{p,n+1}  \right] \label{bJ}
 \bigr\} f_{pq},
\end{align}
where we have defined
\begin{subequations}\label{B&C}
\begin{align}
B_{np} = & \,{1}/{(k w_0)}  \left( \sqrt{p} \, \delta_{p,n+1} - \sqrt{n} \, \delta_{n,p+1}\right), \label{Bnp} \\
C_{np} = & \, {(k w_0)}  \left( \sqrt{p} \, \delta_{p,n+1} + \sqrt{n} \, \delta_{n,p+1}\right)/2. \label{Cnp}
\end{align}
\end{subequations}
From the expressions above we note that both $\bP$ and $\bJ$ do not depend upon $z$. However, Eqs. (\ref{Jperp}) have general validity and they must be satisfied for all values of $z$. This implies that  $\langle \vvr_\perp \rangle$  must be a linear function of $z$, as it can be easily seen by taking the derivative with respect to $z$ of both sides of both Eqs. (\ref{Jperp}),  obtaining
\begin{align}\label{Konst}
{\di \langle \vvr_\perp \rangle}/{\di z} = {\bP_\perp}/{P_z}.
\end{align}
This equation simply states that the centroid of the beam propagates along the axis $z$ obeying the laws of geometrical optics.

To enlighten our results, it is instructive to consider the  specific function $f(\vvr) = f_{00} \psi_{00}+f_{01} \psi_{01}+f_{10} \psi_{10}$ that encompasses several interesting cases. For example, if  $f_{00}=1, \, f_{01} = 0 = f_{10}$  we have the fundamental Gaussian beam with $l=0$ OAM, while for $f_{00}=0, \, f_{01} = i/\sqrt{2}, \, f_{10} = 1/\sqrt{2} $, $f$ reproduces a Laguerre-Gaussian beam $\mathrm{LG}^1_0$ with OAM $l = 1$ \cite{Allen92}. From Eq. (\ref{bP}) it immediately follows
\begin{align}\label{Pt}
P_x = \theta_0 \, \mathrm{Im} \left( f_{00}^* f_{10}\right), \quad P_y = \theta_0 \, \mathrm{Im} \left( f_{00}^* f_{01}\right),
\end{align}
where $\theta_0 = 2/(k w_0)$ is the angular spread of the beam, and $P_z =\abs{f_{00}}^2 + \abs{f_{01}}^2  + \abs{f_{10}}^2$. For sake of simplicity, we  normalize $P_z =1$.
If $f_{00} \in \mathbb{R}$, Eqs. (\ref{Pt}) show that either $f_{10}$ or $f_{01}$ must have an \emph{imaginary} part to guarantee  that the axis of propagation of the beam is tilted with respect to the axis $z$ \cite{TrepsTilt}. Similarly, from Eq. (\ref{bJ}) we obtain
\begin{subequations}\label{J0}
\begin{align}
J_x  = &   \, w_0 \,{\text{Re} \left( f_{00}^* f_{10} \right)}, \label{Jx0} \\
J_y  = &  \, w_0 \,{\text{Re} \left( f_{00}^* f_{01} \right)}, \label{Jy0} \\
J_z  = & \, \lbar \sigma +  2\,\lbar \,  \text{Im} \left( f_{10}^* f_{01}\right). \label{Jz0}
\end{align}
\end{subequations}
Again, if $f_{00} \in \mathbb{R}$, Eqs. (\ref{Jx0}) and (\ref{Jy0}) imply that in order to have a nonzero transverse orbital angular momentum either $f_{10}$ or $f_{01}$ must have a \emph{real} part. It is  known that a superposition with real coefficients of the fundamental mode $\psi_{00}$ with either $\psi_{10}$ or $\psi_{01}$  describes approximatively a \emph{displaced} Gaussian beam  \cite{TrepsTilt}. Thus, we have shown that a lateral displacement of a Gaussian beam changes its transverse OAM. On the other hand, it is also known that such a displacement does not affect the longitudinal OAM $J_z$ \cite{BerrySPIE,VasneTilt}, as it is confirmed by Eq. (\ref{Jz0}) whose orbital part goes to zero when both $f_{10}$ and $f_{01}$ are real numbers.
However, for a \emph{pure} Laguerre-Gaussian beam $\mathrm{LG}^1_0$ one has $f_{00} =0$ and $\bJ_\perp =0$.
Moreover,  in this case $2 \, \text{Im} \left( f_{10}^* f_{01}\right) = 1$ and Eq. (\ref{Jz0}) furnishes $J_z = \lbar \left(\sigma + 1\right)$, in agreement with previous calculations \cite{Allen92}.
 \vspace{-0.22cm}
{\flushleft\emph{\hspace{.3cm}Tilted beams.$\,$}\rule[1.8pt]{0.3cm}{.4pt}}
Equation (\ref{bJ}) clearly shows that $\bJ_\perp$ does not depend from the spin $\sigma$ for any paraxial beam. Another simple way to see this is to substitute Eq. (\ref{rel1}) into  Eqs. (\ref{jx}) and (\ref{jy}). However, such conclusion is not consistent with the following argument: Consider a circularly polarized Gaussian beam $\psi_{00}$ that propagates along the axis $z'$ tilted by an angle $\theta$ with respect to the reference axis $z$. In the Cartesian frame $K'$ attached to the beam, there is a unit of spin angular momentum directed along  $z'$: $j_{z'} = \lbar \, {\sigma} $, and $\bj_{\perp'}=0$.  Differently, in the Cartesian frame $S$ attached to the reference axis $z$, the angular momentum of the beam will have both a longitudinal and a transverse component equal to $j_z = \lbar \, \sigma \cos \theta$,  and $\abs{\bj_\perp} = \lbar \, \abs{\sigma} \sin \theta$, respectively. As the cross-section of the beam when seen from $K$ is augmented by a factor $1/\cos \theta$, we expect that the transverse angular momentum of the tilted beam will go like $\bJ_\perp \sim {\lbar} \, \sigma \tan \theta$ in contrast with Eq. (\ref{bJ}).
Innocent as it is, this conclusion has a striking consequence.  In fact, from Eq. (\ref{Jperp0}) it follows that for our tilted beam $\left.\abs{\langle \vvr_\perp \rangle\right|_{z=0}} \propto \lbar \, \abs{\sigma} \tan \theta $ which is either equal to zero or to $\lbar \, \tan \theta$ when the beam is either linearly or circularly polarized. This means that the position of the barycenter of a Gaussian beam observed from a reference frame non collinear with the direction of propagation of the beam, changes according to the polarization of the beam!
It is worth noting that since the two non-collinear axes $z$ and $z'$ defines uniquely a ``plane of incidence'', then Eq. (\ref{ratioJ}) shows that the barycenter of the tilted beam is shifted in a direction orthogonal to such plane of incidence. In this respect, this effect resembles the so called spin Hall effect of light (SHEL) that was recently measured \cite{HostenandKwiat}.

Now we will support the freshman argument presented above with rigorous calculations.
First, we parameterize the axis of propagation $\hvz'$ of the beam as  $\hvz' = \hvx \sin \theta \cos \phi + \hvy \sin \theta \sin \phi + \hvz \cos \theta  \equiv R(\theta,\phi) \hvz $, where $R(\theta,\phi)= \exp \left( \theta \hvn \cdot \mathbf{L} \right)$ denotes the rotation matrix that connects $K$ with $K'$. Here  $\hvn = (\hvz \times \hvz')/(\abs{\hvz \times \hvz'})$ is a unit vector perpendicular to the plane of incidence and $\mathbf{L}$ is a set of three matrices $\{ L_1,L_2,L_3 \}$ whose component are $[L_i]_{jk} = - \epsilon_{ijk}$, where $\epsilon_{ijk}$ is the Levi-Civita tensor.
Next, we choose   $f = \psi_{00}(\vvr')$ as  solution of the paraxial wave equation in the the beam frame $K'$. We substitute $\psi_{00}(\vvr')$ into Eqs. (\ref{p}) to arrive at the following expression for the energy density flux in $K'$:
\begin{align}\nonumber
\bp' (\vvr')  \propto & \frac{ e^{-L \frac{{x'}^2 + {y'}^2}{ {z'}^2 + L^2} }}{\left( {z'}^2  +L^2\right)^2} \Bigl[ \hvx' \left( {x'} {z'} - {y'} \sigma  L \right) + \hvy' \left( {y'} {z'} +  {x'} \sigma L \right) \Bigr. \\  \label{pK1}
&  \hphantom{xxxxiixxx} \Bigl.  \, + \hvz' \bigr( {z'}^2 + L^2 \bigl) \Bigr].
\end{align}
Finally, we transform $\bp' (\vvr')$ to the frame $K$ via the  map
\begin{equation}\label{pK}
\bp' (\vvr')  \longrightarrow \bp (\vvr) = R(\theta,\phi)  \bp \bigl(R^{-1}(\theta,\phi)\vvr \bigr).
\end{equation}
The resulting expression for $\bp (\vvr)$ is very cumbersome and it will not be reported here. However, as $J_x$ and $J_y$ equals $\langle y \rangle$ and $-\langle x \rangle$ respectively, at $z=0$ only, we can specialize to this case. Moreover, if we consider independently the two cases of horizontal ($\phi =0$), and vertical tilt  ($\phi =\pi/2$) of the beam, we arrive at the following simple expressions for the density of the energy flow $p(x,y,z)$ in the $z = 0$ plane:
\begin{align}\label{fi0}
p_z(x,0,0)  \propto  &  \, e^{-2 \xi^2} \left( 1 - \theta_0 \xi \tan \theta \right),  \\
p_z(0,y,0)  \propto &   \, e^{-2 \eta^2} \left( 1 + \theta_0 \eta \tan \theta \right) ,  \label{fi90}
\end{align}
were, for clarity, we defined the dimensionless variables $\xi = x/w_0$ and  $\eta = y/w_0$. For a well collimated beam $\theta_0 \ll 1$ and from Eqs. (\ref{fi90}-\ref{fi0}) it immediately follows
\begin{align}\label{xAv}
\langle x \rangle = &   \, - \lbar ( \sigma  /2) \tan \theta ,  \\  \label{yAv}
\langle y \rangle = &  \, \lbar  ( \sigma  /2) \tan \theta ,
\end{align}
in agreement with our qualitative argument. For arbitrary values of $z$ and $\phi$ the equations above generalize to
\begin{subequations}\label{rz}
\begin{align}
\langle x \rangle  = &   \, - \lbar \left(\sigma /2 \right) \tan \theta \sin \phi + z \tan \theta \cos \phi, \label{x0z} \\
\langle y \rangle   = &  \,  \lbar \left(\sigma /2 \right) \tan \theta \cos \phi + z \tan \theta \sin \phi . \label{y0z}
\end{align}
\end{subequations}
As expected for a tilted beam, $\di \langle \vvr_\perp \rangle/ \di z = \text{const.}$, namely $\bP_\perp = \text{const.}$ This intuitive result can be verified by a straightforward calculation that produces
\begin{align}
\bP/P_z = & \hvx \tan \theta \cos \phi + \hvy \tan \theta \sin \phi + \hvz,
\end{align}
in agreement with Eqs. (\ref{rz}) and (\ref{Konst}). Finally, we calculate the angular momentum per unit length obtaining
\begin{subequations}\label{j00}
\begin{align}
J_x/P_z  = &   \, \lbar \left(\sigma /2 \right) \tan \theta \cos \phi , \label{jx00} \\
J_y/P_z  = &  \, \lbar  \left(\sigma /2 \right) \tan \theta \sin \phi , \label{jy00} \\
J_z/P_z  = & \, \lbar \left(\sigma /2 \right)\left( 2 - \sin^2  \theta \right)\sec^2 \theta. \label{jz00}
\end{align}
\end{subequations}
It is worth nothing that $\left( J_x^2 + J_y^2 + J_z^2 \right)^{1/2}/P_z = (\sigma/2)(4 -\sin^2 \theta)^{1/2}\sec^2 \theta $, that is $\bJ$ is \emph{not} conserved by the rotation. This is not strange because only the magnitude of the spin density $\bj$ must be invariant under rotation. In fact, as the position vector $\vvr$ in Eq. (\ref{angular}) is defined in the reference frame $K$ and, therefore, is not subjected to the rotation, then the resultant $\bJ$ is frame-dependent and cannot be invariant. Finally, it is easy to check that Eqs. (\ref{Jperp}) are automatically satisfied by the quantities expressed in Eqs. (\ref{rz}-\ref{j00}).
 \vspace{-0.22cm}
{\flushleft\emph{\hspace{.3cm}Conclusions.$\,$}\rule[1.8pt]{0.3cm}{.4pt}}
We have presented a quite counterintuitive and intriguing phenomenon that occurs with tilted beams of light: The position of the intensity barycenter of a tilted Gaussian beam, varies with the polarization of the beam itself. The essence of the phenomenon is that if the beam possesses a \emph{transverse} angular momentum, necessarily the barycenter of its intensity distribution is shifted with respect to its axis of propagation (axis identified with the direction of the Poynting vector). Such a shift occurs in a direction orthogonal to the transverse part of the angular momentum of the beam. This phenomenon shares many  features with the spin Hall effect of light, but it is \emph{not} generated by the interaction between light and matter. For this reason it could be referred to as the ``geometric spin Hall effect of light''. Experiments are in progress in our labs to confirm the existence of this curious phenomenon.
%
%
%

\end{document}